\documentclass{epl}

\title{Finite temperature Functional RG, droplets and decaying Burgers Turbulence}
\author{Pierre Le Doussal}
\institute{CNRS-Laboratoire de Physique Th\'eorique de l'Ecole Normale Sup\'erieure,
24 Rue Lhomond 75231 Paris,
France}

\begin{document}

\maketitle

\newcommand{\fig}[2]{\includegraphics[width=#1]{#2}}

\date{\today}

\begin{abstract}
The functional RG (FRG) approach to pinning of $d$-dimensional
manifolds is reexamined at any temperature $T$. A simple relation
between the coupling function $R(u)$ and a physical observable is
shown in any $d$. In $d=0$ its beta function is displayed to a high
order, ambiguities resolved; for random field disorder (Sinai model)
we obtain exactly the $T=0$ fixed point $R(u)$ as well as its
thermal boundary layer (TBL) form (i.e. for $u \sim T$) at $T>0$.
Connection between FRG in $d=0$ and decaying Burgers is discussed.
An exact solution to the functional RG hierarchy in the TBL is
obtained for any $d$ and related to droplet probabilities.
\end{abstract}

Elastic manifolds pinned by quenched disorder
\cite{pinning} are the simplest
system to study glass phases where (dimensionless) temperature is
formally irrelevant, scaling as $\tilde T_L = T L^{- \theta}$ with
system size $L$. They are parameterized by a displacement
($N$-component height) field $u(x) \equiv u_x$, where $x$ spans a
$d$-dimensional internal space. The competition between elasticity
and disorder produces rough ground states with sample averages $\overline{(u_x-u_{x'})^2}
\sim |x-x'|^{2 \zeta}$ ($\theta=d-2+2\zeta$). These are believed to be statistically scale
invariant, hence should be described by a critical (continuum) field
theory (FT). The latter seems highly unconventional in several
respects. First, an infinite number of operators become marginal
simultaneously in $d=4-\epsilon$. This is handled via Functional RG
methods where the relevant coupling constant becomes a function of
the field, $R(u)$, interpreted as the (second cumulant) disorder
correlator \cite{DSFisher1986}. A more formidable difficulty then
arises: at $T=0$ both $R(u)$ and, more generally, the full effective
action functional $\Gamma[u]$, appear to be non-analytic
\footnote{the non-analyticity of $\Gamma$ occurs at {\it finite scale}
(the Larkin length) contrarily to e.g. the critical $\phi^4$ theory} 
around $u=0$. A linear cusp in $R''(u)$
was found in one loop and large $N$ calculations
\cite{DSFisher1986,frgN}.
Qualitative (two mode minimization) and mean field arguments relate
this cusp to multiple metastable states and shock type singularities
in the energy landscape \cite{BalentsBouchaudMezard1996}. As a
consequence, ambiguities arise in loop corrections
\cite{ChauveLeDoussalWiese2000a}. Although candidate renormalizable
FTs have been identified \cite{ChauveLeDoussalWiese2000a,exactRG}
(working directly at $T=0$) this problem has, until now, hampered
derivation of the field theory from first principles (with the
notable exception of the $N=1$ depinning transition
\cite{LeDoussalWieseChauve2002}).

Working at non-zero temperature $T>0$ should help define the theory,
and $\Gamma[u]$ has been argued to remain smooth within a ''thermal
boundary layer'' (TBL) of width $u \sim \tilde T_L$ around $u=0$.
This width however shrinks as $\tilde T_L \to 0$ in the
thermodynamic limit, and if a fixed $u$, large $L$ limit exists for
any fixed small $T$ it should unambiguously define the
(non-analytic) ``zero temperature theory'': this program, called
''matching'', was proposed and extensively studied in Ref.\cite{BalentsLeDoussal2004a}. 
It does, to some extent, rely on
a scaling ansatz proposed there for the TBL. This ansatz was shown
to be consistent to 1-loop with the droplet picture
\cite{BalentsLeDoussal2004a} and, in the (near) equilibrium (driven)
dynamics, to account for the phenomenology of ultraslow activated
(creep) motion \cite{creep,BalentsLeDoussal2004a}. Although its
physics is reasonable, it is not yet established how a critical
renormalizable FT emerges from it as $\tilde T_L \to 0$, with a
finite unambiguous beta function.

Another field of physics where an (unconventional) field theoretic
description is needed, but remains elusive, is high Reynolds number
turbulence. There too the scale invariant regime, the inertial
range, needs regularization at the small dissipation scale set by
the (formally irrelevant) viscosity $\nu$ \cite{turbBernard}.
Connections between these two tantalizing problems can be made
quantitative within the simplified Burgers turbulence, a much
studied problem \cite{BernardGawedzki98,BouchaudMezardParisi95,Polyakov95,HeinanE}.

Given its central role in the FRG, it is of high interest to obtain
the {\it precise} physical content of the (fixed point) function
$R(u)$, beyond previous qualitative arguments. In the FT, a precise,
but abstract, definition was given, from the replicated effective
action at zero momentum, which allowed for a systematic dimensional
expansion. From it, it was observed that $R''(0)$ gives the exact
sample to sample variance of the center of mass of the manifold (a
typical observable with a universal $T=0$ limit), while $R''''(0)$
yields sample to sample susceptibility fluctuations (a finite
temperature observable which diverges as $T_L \to 0$). It would be
useful to relate directly the full function $R(u)$ to an observable
and cleanly separate zero from finite $T$ contributions.

The present Letter is a short account of a recent study \cite{pld}
aimed at clarifying the physics encoded in the FRG and its
connections to Burgers turbulence. We obtain a simple operational
definition valid in any $d$, not only for $R(u)$, but also for
higher cumulants, and the full (replicated) effective action
$\Gamma[u]$. It makes explicit its $T=0$ physics and at $T>0$ makes
precise the relation between the TBL form of the effective action,
droplet probabilities, and dilute (functional) shocks, via a
(functional) decaying Burgers equation. Next, the instructive
$d=0$ case is studied. For $N=1$, the
matching program started in Ref. \cite{BalentsLeDoussal2004a} is
pushed to obtain here the (unambiguous) beta function to four loop,
and related to works on the inviscid distributional limit
of Burgers equation \cite{BernardGawedzki98}. In the sub-case
of the Sinai (i.e. random field) model, the exact $R(u)$ is computed
at $T=0$. The TBL rounding form at $T>0$ is also obtained.
Obtaining the thermal rounding form in any $d$ amounts to solve an
infinite hierarchy of (functional) exact RG equations: remarkably,
this can be achieved, the solution being parameterized by droplet
probability data. All details are given in \cite{pld}.

The model studied here is defined by the total energy:
\begin{eqnarray}
&& H_V[u] = \frac{1}{2} \int_{xy} g^{-1}_{xy} u_x u_y + \int_x V(u_x,x)
\label{model}
\end{eqnarray}
in a given sample ($u \in R^N$). The distribution of the random
potential is translationally invariant, with second cumulant $
\overline{V(u,x) V(u',x')} = \delta^d(x-x') R_0(u-u')$ and
$\overline{V(u,x)}=0$. This implies the statistical (tilt) symmetry
(STS) under $(x,u_x) \to (x,u_x + \phi_x)$ \cite{pinning}. Several
results here are valid for arbitrary $g_{xy}$, but we often
specialize to $g^{-1}_q=q^2 + m^2$ in Fourier space, where the small
mass provides a confining parabolic potential and a convenient
infrared cutoff at large scale $L_m = 1/m$. In all formula below one
can replace $\int_x  \equiv \int d^dx \to \sum_x$ and $\delta(x-x') \to \delta_{xx'}$ in the 
bare disorder correlator, i.e. a lattice
provides a UV cutoff which preserves STS 
Numerous physical systems are modelled by (\ref{model}) e.g.: (i)
CDW or vortex lattices
\cite{pinning}, for a periodic
$R_0(u)$, (ii) magnetic interfaces with bond disorder, for a short
range (SR) $R_0(u)$, (iii) magnetic interfaces with random field
disorder, of variance $\sigma$, for a long range $R_0(u) \sim -
\sigma |u|$ at large $u$.

Let us briefly recall the convenient definition of $R(u)$ used in
the FT. The model is studied using replica fields $u^a_x$,
$a=1,..,p$, with bare action:
\begin{equation}
{\cal S}[u] = \frac{1}{2 T} \sum_a \int_{xy} g^{-1}_{xy} u^a_x u^a_y
- \frac{1}{2 T^2} \sum_{ab} \int_{x} R_0(u^a_x - u^b_x)
\label{modeld}
\end{equation}
and disorder-averaged correlations of (\ref{model}) identify with
replica correlation functions of (\ref{modeld}) at $p=0$. These (the
connected ones) are obtained from Taylor expanding the $W$
functional, $W[j]= \ln \int \prod_{a x} d u^a_x e^{ \int_{x} \sum_a
j^a_x u^a_x - {\cal S}[u] }$. From it one defines, via a Legendre
Transform, the effective action of the replica theory, $\Gamma[u] =
\int_{x} \sum_a u^a_x j^a_x - W[j]$. It generates (in a Taylor
expansion in $u$) the renormalized vertices, i.e. those where loops
have been integrated, and is thus the important functional for the
FRG. To define the renormalized disorder one assumes an expansion in
number of replica sums:
\begin{equation}
 \Gamma[u] = \sum_{a} \int_{xy} \frac{g^{-1}_{xy} u^a_x
u^a_y}{2 T} - \sum_{ab} \frac{R[u^{ab}]}{2 T^2} - \sum_{abc}
\frac{S[u^{abc}]}{3! T^3} + \cdots \label{gammaud}
\end{equation}
where STS implies that the single-replica term is the bare one, and
the form of the $n$-replica terms, e.g. $R[u^{ab}]$ is a functional
depending only on the field $u^{ab}_x \equiv u^a_x-u^b_x$, whose
value for a uniform field (i.e. local part) defines $R(u)$, i.e
$R[\{u^{ab}_x=u \}]=L^d R(u)$ \footnote{such zero momentum
renormalization conditions are standard in a massive theory. It is
not presently known how to close FRG using other conditions, e.g.
symmetric external momenta, as in massless theories}. It was used in
the FT \cite{ChauveLeDoussalWiese2000a,exactRG} to compute the beta
function, $-m
\partial_m|_{R_0} R(u)=\beta[R](u)$, in powers of $R$,
and its derivatives.

We have shown that this abstract definition is equivalent to a physical one:
for each realization of the random potential $V$, one defines the
renormalized potential functional $\hat V[v]=\hat V[\{v_x\}]$ as the
free energy of the system when centering the quadratic potential
around $u_x=v_x$:
\begin{eqnarray}
&& \hat V[\{v_x\}] = - T \ln \int \prod_{x} d u_x \exp( - H_{V,v}[u]/T ) \label{def}\\
&& H_{V,v}[u] = \frac{1}{2} \int_{xy} g^{-1}_{xy} (u_x-v_x)
(u_y-v_y) + \int_{x} V(u_x,x) \nonumber
\end{eqnarray}
Using STS one sees that the renormalized energy landscape
has second cumulant correlations:
\begin{eqnarray}
&& \overline{ \hat V[\{v_x\}] \hat V[\{v'_x\}] } = \hat
R[\{v_x-v_x'\}]
\end{eqnarray}
and $\overline{\hat V}=0$ (averages are w.r.t. $V$). The result
shown in \cite{pld} is that $\hat R=R$. Hence one can {\it measure}
the 2-replica part of the effective action by computing the free
energy in a well whose position is varied. Choosing a uniform $v_x=v$,
one obtains its local part:
\begin{eqnarray}
&& \overline{ \hat V(v) \hat V(v') } = L^d R(v-v') \label{ru}
\end{eqnarray} where $\hat V(v)=\hat V[\{v_x = v \}]$, using a
parabolic potential centered at $u_x=v$. Performing the Legendre
transform \cite{pld} (more involved) relations are found for
higher cumulants, e.g. $S=\hat S - 3 sym_{abc} g_{xy}
R'_x[v^{ab}] R'_y[v^{ac}]$. The STS property was used: for a non STS
model, e.g. with discrete $u$, either it flows to the
STS fixed point as $m \to 0$, and the above holds asymptotically, or
it does not and a (more involved) extension holds \cite{pld}.

From (\ref{def}) the {\it renormalized (pinning) force} functional
${\cal F}_x=\hat V'_x[v] \equiv \delta \hat V[v]/\delta v_x$ is related to the
thermally averaged position in presence of the shifted well, via
${\cal F}_x= \int_y g^{-1}_{xy} (v_y - \langle u_y
\rangle_{H_{V,v}})$. Hence the force correlator functional
$R''_{xy}[v]$ has a nice expression. For uniform $v_x=v$ and
at $T=0$ it is simple: denote $u_x(v)$ the minimum energy
configuration of $H_{V,v}[u]$ for a fixed $v_x=v$ and $\bar
u(v)=L^{-d} \int_x u_x(v)$ its center of mass position. Then,
denoting $\Delta(v)=-R''(v)$:
\begin{eqnarray}
&& \overline{(v-\bar u(v))(v'-\bar u(v'))} = \Delta(v-v') L^{-d}
m^{-4} \label{delta}
\end{eqnarray}
which generalizes to non-zero $T$ (replacing $\bar u(v)$ by its
thermal average) and to the full multi-local functional
\footnote{(\ref{ru}), (\ref{delta}) generalize to any $N$,
and to two copies as used in chaos studies \cite{chaospld}
$\overline{ \hat V_i(v) \hat V_j(v') } =
L^d R_{ij}(v-v')$. }

For
fixed $L/a$, where $a$ is the UV cutoff scale, the minimum is expected
unique for continuous distributions of $V$, except for a discrete
set of values $v_s$ which are positions of shocks where $\bar u(v)$
switches between different values (e.g. $u_{1}$ to $u_2$) and the
force is discontinuous (at $T=0$): below, the strength of each shock
is noted  $u_{21}^{(s)}=u_2-u_1$.

The renormalized pinning force satisfies an exact RG (ERG) equation
(with $\partial g =-m\partial_m g$):
\begin{equation}
- 2 m \partial_m {\cal F}_x[v] = \int_{yz} \partial g_{yz} ( T {\cal
F}''_{xyz}[v] - {\cal F}'_{xy}[v] {\cal F}_z[v]) \label{burgersf}
\end{equation}
a functional generalization of the {\it decaying Burgers equation}
to which it reduces for $d=0$:
\begin{equation}
 \partial_t F(v) = \frac{T}{2} F''(v) - F'(v) F(v)
\label{burgers}
\end{equation}
with $t=m^{-2}$, $F(v)=\hat V'(v)$, usually written $\partial_t {\sf
u} + {\sf u}'_x {\sf u} = \nu {\sf u}''_{xx}$, identifying ${\sf
u},x,\nu$ in Burgers to $F,v,T/2$ in the FRG (while $\hat V[v]$
satisfies a functional KPZ-type equation). The stochasticity in
(\ref{burgersf}),(\ref{burgers}) comes from their (random) initial
conditions $F(v)=V'(v)$ and ${\cal F}_x[v]=V'_x[v]$ for
$t=0,m=\infty$. Eq. (\ref{burgers}) (and its primitive) is
equivalent to an infinite ERG hierarchy for the n-th moments $\bar
S^{(n)}(v_{1,2,..,n})=(-)^n \overline{\hat V(v_1)..\hat V(v_n)}$ in
$d=0$:
\begin{eqnarray}
&& - m \partial_m R(v) = \frac{2 T}{m^2} R''(v) + \frac{2}{m^2} \bar
S_{110}(0,0,v) \label{h0}  \\
&& - m \partial_m \bar S^{(n)}(v_{1,2,..,n}) = \frac{n T}{m^2} [
\bar S^{(n)}_{20..0}(v_{1,2,..n})] + \frac{n}{m^2} [\bar S^{(n+1)}_{110..0}(v_{1,1,2..n})]
\label{hierarchy0}
\end{eqnarray}
where $\bar S \equiv \bar S^{(3)}=\hat S$, subscripts denote partial
derivatives and $[..]$ is symmetrization. A similar, more formidable
looking functional hierarchy exists for any $d$:
\begin{equation}
 - m \partial_m  R[v] = T \partial g_{xy} R''_{xy}[v] +
\partial g_{zz'} \bar S^{110}_{zz'}[0,0,v] \label{ergd}
\end{equation}
together with ERG equations for $\bar S$ and higher moments. In both
cases a related hierarchy exists for the cumulants $R$, $S$, .. defining
$\Gamma[u]$ in (\ref{gammaud}), studied in
\cite{BalentsLeDoussal2004a,schehr_co_pre}. The usual RG strategy is
to truncate them to a given order in $R$ yielding the beta function.
Ambiguities in the limit of coinciding arguments in
(\ref{hierarchy0},\ref{ergd}) may arise in doing so directly at
$T=0$.

We start with $d=0$ (and $N=1$), a particle in a 1D random potential
$V(u)$, aiming to obtain an unambiguous beta function as $m \to 0$.
We define rescaled $\tilde T = 2 T m^\theta$ and $R(u) = \frac{1}{4} m^{\epsilon-4 \zeta} \tilde
R(u m^{\zeta})$ [this should yield a
FP when correlations of $V$ grow as $u^{\theta/\zeta}$]. Trying
first standard loop expansion at $T>0$ ($\tilde R$ analytic), we
obtained from (\ref{hierarchy0}) the beta function $- m
\partial_m \tilde R|_{R_0} = \beta[\tilde R,T]$. To $n$ loop, it is
a sum of terms of order $\tilde T^p \tilde R^{n+1-p}$, $0 \leq p
\leq n$. The one-loop equation (i.e. adding $\tilde T \tilde R''$ to
the first three terms in (\ref{rg4b}) below) exhibits the standard TBL
for $u \sim \tilde T$ discussed in \cite{BalentsLeDoussal2004a}. To
2-loop a term $\frac{-1}{4} \tilde T \tilde R''''(0) \tilde
R''(u)$ appears, and using the TBL identity $\lim_{m \to 0} \tilde T \tilde
R''''(0) = \tilde R'''(0^+)^2$, exact at one loop, produces
precisely the 2-loop ``anomalous'' term in (\ref{rg4b}) below. Alas, one finds \cite{pld} that this
procedure fails at 3-loop. One must instead examine the whole ERG hierarchy
as in \cite{BalentsLeDoussal2004a}. There, a method to obtain the
unambiguous beta function was found by verifying order
by order, a continuity property of the $\Gamma$-cumulants $S^{(n)}_{11..1}(u_{1..n})$ upon bringing points together. We completed in \cite{pld} the derivation of the
(local) beta function, obtaining (up to a constant, with ${\sf R}''=\tilde R''-\tilde R''(0)$):
%\begin{widetext}
\begin{eqnarray}
&& -m \partial_m \tilde R = (\epsilon - 4 \zeta) \tilde R + \zeta u
\tilde R' + [ \frac{1}{2} (\tilde R'')^2 - \tilde R''(0) \tilde
R'' ] + \frac{1}{4} ((\tilde R''')^2- \tilde R'''(0^+)^2 ) {\sf R}'' \label{rg4b} \\
&& + \frac{1}{16} ({\sf R}'')^2 (\tilde R'''')^2 + \frac{3}{32}
((\tilde R''')^2- \tilde R'''(0^+)^2)^2 + \frac{1}{4} {\sf R}''
((\tilde R''')^2 \tilde R''''- \tilde R'''(0^+)^2 \tilde
R''''(0^+)) \nonumber \\
&& + \frac{1}{96} ({\sf R}'')^3 (\tilde R^{(5)})^2 + \frac{3}{16} ({\sf
R}'')^2 \tilde R''' \tilde R'''' \tilde R^{(5)} + \frac{1}{8}
{\sf R}'' ((\tilde R''')^3 \tilde R^{(5)} - \tilde R'''(0^+)^3 \tilde R^{(5)}(0^+)) \nonumber  \\
&&  + \frac{1}{16} ({\sf R}'')^2 (\tilde R'''')^3 
+ \frac{9}{16} {\sf R}'' ( (\tilde R''')^2 (\tilde R'''')^2 - \frac{1}{6}
R'''(0^+)^2 (\tilde R'''')^2 - \frac{5}{6} \tilde R'''(0^+)^2 \tilde R''''(0^+)^2 ) \nonumber  \\
&& + \frac{5}{16} ((\tilde R''')^2-\tilde R'''(0^+)^2) ((\tilde R''')^2 \tilde R'''' +
\frac{1}{10} \tilde R'''' \tilde R'''(0^+)^2 - \frac{11}{10} \tilde R''''(0^+)
\tilde R'''(0^+)^2 ) + O(\tilde R^6) \nonumber
\end{eqnarray}
%\end{widetext}
The first line are one and
2-loop terms, the second is 3-loop, the last three are 4-loop. 
Normal terms (i.e. non vanishing for analytic
$R(u)$) are grouped with anomalous ''counterparts'' to show the
absence of $O(u)$ term, a strong constraint (linear cusp, no
supercusp): these combinations can hardly be guessed beyond 3 loop.
This shows the difficulty in constructing the FT, already in $d=0$.
We emphasize that (\ref{rg4b}) results from {\it a first principle derivation}.

\begin{figure}
\centerline{\includegraphics[width=5cm]{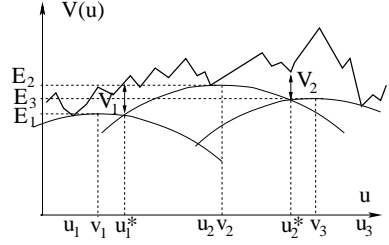}} \caption{
Construction of the joint probability $P(\{E_i,v_i\})$ that $\hat
V(v_i)=E_i$ at points $v_i$: the random walk $V(u)$ must remain
above all parabola centered on the $v_i$ of apex $E_i$ intersecting at
points $u_i^*$. Each
independent interval $[u_i^*,u_{i+1}^*]$ can be treated as in
\cite{LeDoussalMonthus2003}. \label{shockR}}
\end{figure}

$R(u)$ being a physical observable, we look for cases where it
can be computed. The Brownian landscape $V(u)$, the so-called Sinai
model, is interesting as the $d=0$ limit of random field disorder.
Recently we obtained the full statistics of (deep) extrema in
presence of a harmonic well \cite{LeDoussalMonthus2003}. This is
generalized \cite{pld} as described in Fig \ref{shockR}. Graphically
the renormalized landscape $\hat V(v) = E$ is constructed by raising
a parabola $P_v$, $y(u) = - \frac{m^2}{2} (u-v)^2 + E'$ from $E' =
-\infty$ until it touches (for $E'=E$) the curve $y=V(u)$ at point
$u=u_1(v)$, position of the minimum of $H_{V,v}(u)$, $E$ being the
maximum (apex) of the parabola. $P_v$ touching at two points
$u_1(v)<u_2(v)$ signals a shock at $u=v$. Computing
$P(E_1,v_1;E_2,v_2)$ (see Fig. \ref{shockR}) yields \cite{pld} (for
$m^2=1$, $\sigma=1$) and $\bar R(v)=R(v)-R(0)$:

%\begin{widetext}
\begin{equation}
\bar R(v) = - 2^{\frac{1}{3}} \sqrt{\pi v} e^{- \frac{v^3}{48}}
\int_{\lambda_1} \int_{\lambda_2} \frac{[1 - \frac{2
(\lambda_2-\lambda_1)^2}{b^2 v}] e^{i \frac{v}{2 b}
(\lambda_1+\lambda_2) - \frac{(\lambda_2-\lambda_1)^2}{b^2 v}}
}{Ai(i \lambda_1) Ai(i \lambda_2)} [1+ \frac{v \int_0^\infty dV
e^{\frac{v}{2} V } Ai(a V + i \lambda_1)
 Ai(a V + i \lambda_2)}{Ai(i \lambda_1) Ai(i \lambda_2)} ]
 \label{ru0}
\end{equation}
%\end{widetext}
where $a=2^{-1/3}$, $b a^2=1$, $\int_{\lambda} \equiv
\int_{-\infty}^{+\infty} \frac{d \lambda}{2 \pi}$ and all integrals
converge well. One finds $\bar R(v)\approx -v+0.810775$ at large
$v$, and recovers \cite{LeDoussalMonthus2003} $\overline{u^2}= -
R''(0) = 1.054238$. Once rescaled, (\ref{ru0}) should be a FP of
(\ref{rg4b}) corresponding to $\zeta=\zeta_{RF}=4/3$.

At non-zero $T$, one reexamines (\ref{def}) taking into account,
within a droplet calculation, the probability density $D(y) dy$ for
two degenerate minima of $V$, spatially separated by $y=u_2-u_1$
($D(y)=D(-y)$). It yields the TBL form (for $v \sim T m^2$):
\begin{eqnarray}
R''(v) = R''(0) + m^4 T \langle y^2 F_2(m^2 y v/T) \rangle_y
\label{rounding}
\end{eqnarray}
with $F_2(z) = \frac{z}{4} \coth \frac{z}{2} - \frac{1}{2}$ and
$\langle .. \rangle_y \equiv \int dy .. D(y)$ is normalized by the
STS identity ${\langle y^2 \rangle_{y}}=2/m^2$. Since $F_2(z) \sim
|z|/4$ at large $z$ (\ref{rounding}) yields consistent matching
between finite $T$ (droplet) quantities in the TBL and the cusp of
the $T=0$ FP for $v=O(1)$, with $R'''(0^+) = \frac{m^4}{2}
\frac{\langle |y|^3 \rangle_{y}}{\langle y^2 \rangle_{y}}$.
(\ref{rounding}) should be more generally valid in $d=0$ (any $N$),
but in the RF case it is known \cite{LeDoussalMonthus2003} that
(setting $m=\sigma=1$) $D(y) = \frac{1}{2}
\int_{\lambda_1,\lambda_2} \frac{Ai'(i \lambda_1) e^{i
(\lambda_1-\lambda_2) |y|/b} }{Ai(i \lambda_1) Ai^2(i \lambda_2)}$
found to be consistent with $R'''(0^+) = 0.901289$
from (\ref{ru0}). Remarkably, (\ref{rounding}) generalizes to higher
moments $\bar S^{(n)}$, with generalized functions $F_n$ explicitly
obtained in \cite{pld} yielding an exact ``droplet'' solution of
the hierarchy (\ref{hierarchy0}) within the TBL.

Since in $d=0$ the FRG (\ref{burgers}) {\it identifies} with
decaying Burgers, we emphasize the correspondence:
\begin{equation}
 - R''(0) \equiv \overline{{\sf u}(x)^2} \quad , \quad \frac{T}{2} R''''(0)
\equiv \nu \overline{(\nabla {\sf u}(x))^2} = \bar \epsilon
\label{corr1}
\end{equation}
(more generally $\overline{{\sf u}(x) {\sf u}(0)} \equiv -R''(x)$),
both have finite limits as $\nu \to 0$. The second is the {\it
dissipative anomaly}: also present in 3D Navier Stokes. In Burgers it
is due to shocks. The (equivalent) finite limit of the l.h.s.
implies a thermal boundary layer in the FRG. Dilute shocks in
Burgers are equivalent to droplets and a TBL in the FRG
where $u_{21}= \equiv {\sf u}(0^+)-{\sf u}(0^-)$, and
(\ref{rounding}) can indeed be recovered from a single shock
solution in Burgers upon averaging over its position \cite{pld}. The
celebrated Kolmogorov law in the inertial range:
\begin{equation}
\frac{1}{2} \bar S_{111}(0,0,u) \sim \bar \epsilon u \equiv
 \frac{1}{12} \overline{({\sf u}(x)-{\sf u}(0))^3 } \sim - \bar \epsilon x
\label{corr2}
\end{equation}
corresponds to the non-analytic behaviour of the third cumulant 
at small argument in the $T=0$ theory. Identical coefficients in
(\ref{corr1}) and (\ref{corr2}) are a consequence of matching across
the TBL (i.e. viscous layer), identifying the second derivative
of (\ref{h0}) at $v=0$ (for $\nu>0$) and $v=0^+$ (for $\nu \to 0$)
i.e $\partial_t R''(0) = T R''''(0) \equiv \partial_t R''(0^+) =
\bar S_{112}(0,0,0^+)$.
Similar relations exist in stirred Burgers (and Navier Stokes)
\cite{turbBernard}: there the dissipation rate $\bar \epsilon$ is
balanced by forcing, instead of scale-invariant time decay of
correlations, but small-scale shock properties should be rather
similar. Closure of hierarchies similar to (\ref{hierarchy0}) was
proposed there \cite{Polyakov95} in terms of an ''operator product expansion''.
Recent studies cast doubt on such simple
closures \cite{BernardGawedzki98}: N=1 decaying
Burgers (and stirred \cite{HeinanE}) can be constructed in the
inviscid limit ($\nu \to 0$) using distributions, e.g. $t
F'(v)=1-\sum_s u_{21}^{(s)} \delta(v-v_s)$. It is shown there that
shock ''form factors'' (i.e. size-distribution) determines small
distance (non-analytic) behaviour of moments of velocity
differences, $\overline{(F(v)-F(0))^{p}} \sim \mu_{p} v
sign(v)^{p+1}$, with $\mu_p=\overline{\sum_s (u_{21}^{(s)})^p
\delta(v-v_s)}$. In the FRG these are equivalent to droplet distributions
as we show \cite{pld} that $\mu_p=\langle |y|^{p+1} \rangle_y/\langle
y^{2} \rangle_y$, e.g. consistent with $R'''(0^+) \equiv
\overline{{\sf u}(0^{+}) \nabla {\sf u}(0)}= \mu_{2}/(2 t^2)$ given
above. The $T=0$ distributional limit of (\ref{burgers})
derived in \cite{BernardGawedzki98} is equivalent to
$\partial_t \hat V(v) = -\frac{1}{2} F(v^+) F(v)$: it validates the
first-principle FRG discussed above yielding (\ref{rg4b}) (the
central property being continuity of all $\bar S_{1..1}$ since
$F(v)$ remains bounded). These considerations should be universal
for dilute shocks, i.e. independent of details of shock
probability distributions. For RF disorder the full distribution of
shock parameters $\{u_{21}^{(s)},v_s \}$ is known exactly
\cite{FrachebourgMartin99}. It is used in
\cite{pld} to obtain from (\ref{delta}) another expression for
$\Delta(v)$ fully consistent with (\ref{ru0}).

Extensions to higher $d$ are studied in \cite{pld}. Let us indicate
here that similar droplet estimates can be performed and yield an
exact solution of the full {\it functional} hierarchy (\ref{ergd})
for all moments. Within the TBL $m^2 v_x/T = O(1)$ the $R[v]$
functional reads:
\begin{equation}
R[v] = \frac{1}{2} v_x R_{xy}''[0] v_y + T^3 \sum_i \langle H_2(
\int_{xy} v_x g^{-1}_{xy} u^{(i)}_{12,y}/T) \rangle_D \label{tbl}
\end{equation}
where $H_2''(z)=F_2(z)$. To find this solution one considers a 
small density (of order 
$T m^\theta$) of well separated ``elementary droplets'', 
i.e. local GS degeneracies
$u^{(i)}_{12,x}=u^{(i)}_{2,x}-u_{1x}$. $\langle .. \rangle_D$
denotes the average over them. Eq. (\ref{tbl}) relates droplet
probabilities to the TBL in the FRG.

To conclude we related FRG functions, e.g. $R(u)$, to 
observables. This allows to compute them in simple cases, and provides a 
method to measure them in numerics and experiments. Their relations to
shocks in energy (or force) landscape was made precise, via a generalized 
Burgers equation. We have shown
how shock form factors and droplet distributions are related to the
FRG functions. Questions such as the extent of universality in the TBL,
how do properties of $N=1$, $d=0$ Burgers extend to functional
shocks (e.g. Kolmogorov law) remain tantalizing but can 
now be adressed.

We warmly thank K. Wiese and L. Balents for enlightening discussions
and long standing collaborations on FRG. We acknowledge support from ANR under program
05-BLAN-0099-01

\bigskip

%\begin{references}

\end{document}